\documentclass[twocolumn,showpacs,preprintnumbers,amsmath,amssymb,showkeys]{revtex4}
\usepackage{graphicx}% Include figure files
\usepackage{dcolumn}% Align table columns on decimal point
\usepackage{bm}% bold math

\begin{document}

\preprint{For comments please contact:je2ccc@gmail.com }

\title{Sub-wavelength Coherent Imaging of a Pure-Phase Object with Thermal Light}

\author{Minghui Zhang}
 \affiliation{Key Laboratory for Quantum Optics and the Center for
Cold Atom Physics of CAS,\\ Shanghai Institute of Optics and Fine
Mechanics, Chinese Academy of Science,\\P. O. Box 800-211,
Shanghai,201800, P. R. China}
\author{Qing Wei}
 \affiliation{Key Laboratory for Quantum Optics and the Center for
Cold Atom Physics of CAS,\\ Shanghai Institute of Optics and Fine
Mechanics, Chinese Academy of Science,\\P. O. Box 800-211,
Shanghai,201800, P. R. China}
 \email{Second.Author@institution.edu}
\author{Xia Shen}
 \affiliation{Key Laboratory for Quantum Optics and the Center for
Cold Atom Physics of CAS,\\ Shanghai Institute of Optics and Fine
Mechanics, Chinese Academy of Science,\\P. O. Box 800-211,
Shanghai,201800, P. R. China}
 \homepage{http://www.Second.institution.edu/~Charlie.Author}
\author{Yongfeng Liu}
 \affiliation{Key Laboratory for Quantum Optics and the Center for
Cold Atom Physics of CAS,\\ Shanghai Institute of Optics and Fine
Mechanics, Chinese Academy of Science,\\P. O. Box 800-211,
Shanghai,201800, P. R. China}
\author{Honglin Liu}
 \affiliation{Key Laboratory for Quantum Optics and the Center for
Cold Atom Physics of CAS,\\ Shanghai Institute of Optics and Fine
Mechanics, Chinese Academy of Science,\\P. O. Box 800-211,
Shanghai,201800, P. R. China}
\author{Yanfeng Bai}
 \affiliation{Key Laboratory for Quantum Optics and the Center for
Cold Atom Physics of CAS,\\ Shanghai Institute of Optics and Fine
Mechanics, Chinese Academy of Science,\\P. O. Box 800-211,
Shanghai,201800, P. R. China}
\author{Shensheng Han}
 \affiliation{Key Laboratory for Quantum Optics and the Center for
Cold Atom Physics of CAS,\\ Shanghai Institute of Optics and Fine
Mechanics, Chinese Academy of Science,\\P. O. Box 800-211,
Shanghai,201800, P. R. China}
\date{\today}

\begin{abstract}
We report, for the first time, the observation of sub-wavelength
coherent image of a pure phase object with thermal light,which
represents an accurate Fourier transform. We demonstrate that
ghost-imaging scheme (GI) retrieves \emph{amplitude} transmittance
knowledge of objects rather than the transmitted \emph{intensities}
as the HBT-type imaging  scheme does.
\end{abstract}


\maketitle

\section{\label{1}Introduction}
In many imaging circumstances, phase information about objects plays
a role as well as or even more important than intensity does, for
example, when the objects are pure-phased, that is, highly
transparent and absorb little light, imaging can not be simply
realized by the transmitted or reflected intensity information of
thermal lights. Although phase distribution about an object can be
retrieved from its Fourier-transform diffraction pattern was firstly
proposed by Sayre \cite{sayre} and dedicated efforts described in
the works like \cite{fienup} demonstrated and developed the
techniques, the efforts seems to be in vain if diffraction imaging
applications were in hard $x$-ray, $\gamma$-ray, or other
wavelengths where no effective lens or/and no coherent source is
available. Recent works \cite{kw1,yshi_epl_2004} reported a new
version of the landmark Hanbury Brown and Twiss (HBT) experiment
\cite{hbt} and gave lens-less Fourier-transform pattern of a Youngs
double slit with thermal sources, but the phase information is not
yet mentioned because the object they used was amplitude-only. In
fact, as we will discuss latterly, the classical HBT-type imaging,
features that the joint detection plane in the optical path passes
through the object, will be invalid for retrieving phase knowledge
about an object. Since middle years of last decades, ghost imaging
(GI) has been enthusiastically
studied\cite{Belinsky,cteich_prl_2001,boyd_prl_2001,lg_prl_2003,
boyd_prl_2004,yshi_prl_2004,lg_prl_2004,ch,yshi_prl_2006},
here, the reason for the term \emph{ghost} used  is that the image
of an object, diffractive or geometrical, would appear as a function
of the position in the path that actually never pass the object, and
this unique feature is regarded as a key difference from classical
HBT-type imaging. Although whether the entangled beams was a
prerequisite once have been hotly
debated,\cite{cteich_prl_2001,boyd_prl_2001,lg_prl_2003,boyd_prl_2004,ch}
it is generally accepted now that classical thermally emitted light
can be used for GI and quantum entangled beams is not a
prerequisite. In this letter, we report, for first time, the
lensless retrieval of sub-wave length coherent imaging of a
pure-phase object, which represents an accurate Fourier transform,
within fresnel diffraction range by using GI scheme. Also, we
experimentally demonstrate that ghost-imaging scheme (GI) retrieves
complex \emph{amplitude} transmittance of objects rather than the
transmitted \emph{intensities} as the HBT-type imaging  scheme does.

We first go over the theory skeleton for two-photon joint-detection
and its related physics behind the lensless Fourier transform
imaging in Section~\ref{2}, then, in Section~\ref{3}, the experiment
setup, including preparation for pure-phase object and pseudothermal
light source  and arrangement for instant intensity detecting and
recording were briefly introduced, after that, the experiment
results were shown in Section~\ref{4} by comparing with both
theoretical anticipation and Fourier transform imaging performed by
$2f$ system. The conclusions and discussions were finally in
Section~\ref{5}.

\section{\label{2}The theoretical basis}

\subsection{\label{A}Joint detection}

\begin{figure}
\centerline{\includegraphics* [bb=70 53 272 154,scale=0.9]
{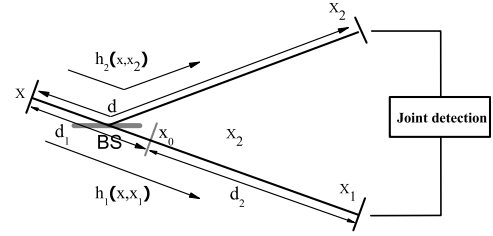}}\caption{\label{fig:setup} Setup schemas for
experiment.$h_{k}(x,x_{k})$ refers to the impulse response function
of both optical paths formed by beam splitter from source plane $X$
to detection plane $X_{k}$;$k=1,2$. }
\end{figure}

Fig.\ref{fig:setup} shows the setup for experiment. Intensity
information is recorded at plane $X_{1}$ and $X_{2}$ in two
different optical paths formed by a $5/5$ none-polarized beam
splitter (BS). $h_{1,2}(x,x_{1,2})$ refers to the impulse response
functions from thermal light plane $X$ to intensity detecting plane
$X_{1,2}$, where, $x$ and $x_{1,2}$ stand for positions on plane $X$
and $X_{1,2}$, respectively.

The physics behind the joint detection for GI scheme in plane
$X_{1}$ and plane $X_{2}$, as Fig.\ref{fig:setup} shows,
can be explained in simplicity as follows: The two-photon amplitude
described by state vector $|A\rangle$ can be decomposed into
weighted sum of the following three normalized basis states:
$|\alpha\rangle=|vac,j\rangle$, two-photons both reflected by BS;
$|\beta\rangle=|m,n\rangle$, one photon reflected by the BS and the
other transmitted the BS; and $|\gamma\rangle=|i,vac\rangle$,
two-photons both transmitted the BS. \emph{i.e.}
$|A\rangle=\frac{1}{2}|\alpha\rangle+\frac{1}{\sqrt{2}}e^{i\theta_{1}}|\beta\rangle+\frac{1}{2}e^{i\theta_{2}}|\gamma\rangle$.
In the equation, $\theta_{1(2)}$ is the  phase of complex weight for
state $|\beta\rangle$($|\gamma\rangle$) relative to state
$|\alpha\rangle$. The expression of $|\alpha\rangle$ and
$|\gamma\rangle$ can be expressed as:
\begin{align}
|\alpha\rangle=|vac\rangle_{1}|j\rangle_{2},\label{eq:1}
\end{align}\\and
\begin{align}
|\gamma\rangle=|i\rangle_{1}|vac\rangle_{2},\label{eq:2}
\end{align}
but as for the characters of identical bosons, the two-photon state
of $|\beta\rangle$ must be expanded in this way:
\begin{align}
|\beta\rangle=\frac{1}{\sqrt{2}}(|m\rangle_{1}|n\rangle_{2}+|n\rangle_{1}|m\rangle_{2}).\label{eq:3}
\end{align}
Among the Eq.(\ref{eq:3}), the subscripts $1$ and $2$ of state
vector $|\rangle$ refer to joint detecting points in plane $X_{1}$
and $X_{2}$ respectively, and $m$, $n$ stand for two
undistinguishable photons.

If we define $E^{(\pm)}(t_{1,2}x_{1,2})$ as the positive-frequency
and negative-frequency components of the field at time-spatial point
$t_{1}x_{1}$ and $t_{2}x_{2}$, suppose time $t_{1}<t_{2}$, the joint
detection probability per unit $(time)^{2}$ that one photon is
recorded at $x_{1}$ at time $t_{1}$ and another at $x_{2}$ at time
$t_{2}$, say, the square module of two-photon amplitude
$\psi(t_{1}x_{1},t_{2}x_{2})$ corresponding to $|A\rangle$ has been
described by second-order Glauber correlation function
\cite{glauber_pr_1963}: $ G^{2}(t_{1}x_{1},t_{2}x_{2})=Tr\{\rho
E^{(-)}(t_{1}x_{1})E^{(-)}(t_{2}x_{2})
E^{(+)}(t_{2}x_{2})E^{(+)}(t_{1}x_{1})\}$, where the density
operator is defined as the average outer product of state
$|A\rangle$: $\rho=\{|A\rangle\langle A|\}_{av}$, \emph{i.e.} ,
\begin{align}
\rho=\frac{1}{4}|\alpha\rangle\langle\alpha|+\frac{1}{2}|\beta\rangle\langle\beta|
+\frac{1}{4}|\gamma\rangle\langle\gamma|.\label{eq:4}
\end{align}
Subsisting Eq.(\ref{eq:1})-Eq.(\ref{eq:4})into second-order Glauber
correlation function, and note that
$E^{(+)}(t_{1,2}x_{1,2})|vac\rangle=\langle
vac|E^{(-)}(t_{1,2}x_{1,2})=0$, we find that only state
$|\beta\rangle$, contributes to the joint detection as:
\begin{align}
G^{(2)}(t_{1}x_{1},t_{2}x_{2})\propto\langle\beta|E^{(-)}(t_{1}x_{1})E^{(-)}(t_{2}x_{2})\nonumber
\\\times
E^{(+)}(t_{2}x_{2})E^{(+)}(t_{1}x_{1})|\beta\rangle.\label{eq:5}
\end{align}

On the other hand, if joint detection is only performed in the same
plane $X_{1}$ in case of HBT-type imaging, which we'll discuss
latterly, we can regard the two-photon amplitude described by state
vector $|A\rangle$ as a pure state only concerning with two photons
both transmitted the beam splitter, and thus the density operator of
state $|A\rangle$ comes to: $\rho=|A\rangle\langle A|$, \emph{i.e.}
and joint detection probability reduces to:
\begin{align}
G^{(2)}(t_{1}x_{1},t_{2}x'_{1})= \langle
A|E^{(-)}(t_{1}x_{1})E^{(-)}(t_{2}x'_{1})\nonumber
\\\times
E^{(+)}(t_{2}x'_{1})E^{(+)}(t_{1}x_{1})|A\rangle,\label{eq:6}
\end{align}
where, $x_{1}$ and $x'_{1}$ stand for two different points in plane
$X_{1}$.

\subsection{\label{B}The relation between joint detection and intensity correlations }
Eq.(\ref{eq:5}) and Eq.(\ref{eq:6}) states the fact that whether the
joint detection performed in both planes $X_{1}$ and $X_{2}$, or in
plane $X_{1}$ only, the joint detection probability would be
presented as average production of normally ordered
positive-frequency and negative-frequency operator. Quantum
detection theory shows that right side of Eq.(\ref{eq:5}) and
Eq.(\ref{eq:6}) is proportional to the normally ordered intensity
correlation function of the optical
field\cite{op_coher_qt_op,qt_op}.\emph{i.e.},
\begin{align}
G^{(2)}(t_{1}x_{1},t_{2}x_{2})\propto\langle
I_{1}(t_{1}x_{1})I_{2}(t_{2}x_{2})\rangle,\label{eq:7}
\end{align}
and
\begin{align}
G^{(2)}(t_{1}x_{1},t_{2}x'_{1})\propto\langle
I_{1}(t_{1}x_{1})I_{2}(t_{2}x'_{1})\rangle.\label{eq:8}
\end{align}
The similarity of Eq.(\ref{eq:7})and Eq.(\ref{eq:8}) means that
square module of two-photon amplitude, can can be measured by
correlation of intensities no matter whether joint detection is
performed in plane $X_{1}$ and $X_{2}$ on two different optical path
formed by beam splitter or only in plane $X_{1}$ on the optical path
transmitted the beam splitter.
\subsection{\label{C}Measuring the modulation to two-photon amplitude}
Previous works\cite{lg_prl_2004,ch} have stated that if one assumes
the time window $\Delta t=t_{2}-t_{1}=0$ in the experiment, and
defines $\Delta I^{(2)}(r_{1}, r_{2})=\langle
I_{1}(r_{1})I_{2}(r_{2})\rangle - \langle I_{1}(r_{1})\rangle
\langle I_{2}(r_{2})\rangle$,  the impulse response functions for
the two optical paths in Fig.\ref{fig:setup}, $h_{1, 2}(x, x_{1,
2})$ would be embedded into:
\begin{align}
\Delta I^{(2)}(x_{1},x_{2})\propto|\int\limits_{X}
h_{1}^{*}(x',x_{1})h_{2}(x,x_{2})\nonumber\\<E^{*}(x')E(x)>dx'dx|^{2}
,\label{eq:9}
\end{align}
with the assumption that the thermal state of light is characterized
by a Gaussian field statistics\cite{st_optics}, among them $x$ and
$x'$ denote position(s) at thermal source plane $X$ in
Fig.\ref{fig:setup}. According to Collins¡¯s integral
formula\cite{collion}, there exists:
\begin{align}
h_{1}(x,x_{1})=\int\limits_{X_{0}}
dx_{0}\frac{e^{-ikd_{1}}}{i\lambda d_{1}}e^{\frac{-i\pi}{\lambda
d_{1}}(x-x_{0})^{2}}\nonumber\\\times
\mathrm{t}(x_{0})\frac{e^{-ikd_{2}}}{i\lambda
d_{2}}e^{\frac{-i\pi}{\lambda d_{2}}(x_{1}-x_{0})^{2}},\label{eq:10}
\end{align}
and
\begin{align}
h_{2}(x,x_{2})=\frac{e^{-ikd}}{\lambda d}e^{\frac{-i\pi}{\lambda
d}(x_{2}-x)^{2}},\label{eq:11}
\end{align}
where, $\mathrm{t}(x_{0})$ is the complex amplitude transmittance of
objects placed on plane $X_{0}$. As for GI schemed imaging, joint
detection was performed on  both optical paths. Under the condition
of $d=d_{1}+d_{2}$, substituting Eq.(\ref{eq:10}) and
Eq.(\ref{eq:11}) into Eq.(\ref{eq:9}), we get:
\begin{align}
\Delta I^{(2)}(x_{1},x_{2})\propto \mid \mathcal F
\{\mathrm{t}[\frac {2\pi(x_{2}-x_{1})}{\lambda
d_{2}}]\}\mid^{2},\label{eq:12}
\end{align}
Among them $\mathcal F\{~\}$ denotes Fourier transform.

In the other case, if joint detection is only performed at plane
$X_{1}$ on the optical path which passes through the object, then
Eq.(\ref{eq:10}) will stand for both of two impulse response
functions in Eq.(9). In this situation, the equation becomes:

\begin{align}
\Delta I^{(2)}(x_{1},x'_{1})\propto \mid \mathcal F
\{|\mathrm{t}\frac {2\pi(x_{1}-x'_{1})}{\lambda
d_{2}}|^{2}\}\mid^{2},\label{13}
\end{align}where, $x_{1}$ and $x'_{1}$ stand for position(s) at
plane $X_{1}$ .

Based on the theory above, the following experiment was carried out
in the regime of large number photons to illustrate the behaviors of
two-photon interference. Now we can easily envision the physics
behind the following experiment: The modulation to the two-photon
amplitude, $\psi(x_{1},x_{2})$ and $\psi(x_{1},x'_{1})$, by the
objects was measured by correlation function of intensity
fluctuations.

\section{\label{3}The experimental setup}
The experiment is set out with the use of pure phase object
\begin{figure}
\centerline{\includegraphics* [bb=29 54 192 127,scale=1.0]
{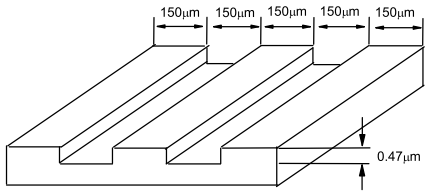}}\caption{\label{fig:obj} The pure-phase object was
prepared by etching two grooves with width of $150\mu m$ and
separating them by a $150\mu m$ un-etched area on a piece of
$0.75mm\times10mm$ square quartz glass (JGS1). The other two
un-etched areas with width of $150\mu m$ are left symmetrically. The
depth of two grooves is arranged to be $0.47\mu m$.}
\end{figure}
, which was prepared by etching two grooves with width of $150\mu m$
and separating them by a $150\mu m$ un-etched area on a piece of
$0.75mm\times10mm$ square quartz glass (JGS1), as Fig.\ref{fig:obj}
shows. The other two un-etched areas with width of $150\mu m$ are
left symmetrically. Since we are using Nd: YAG laser with the
wavelength of $\lambda=0.532\mu m$ , the depth of two grooves is
arranged to be $\lambda/2(n-1)=0.532/2(1.57-1)=0.47\mu m$ to form
phase differences of $\Delta\phi=\pi$ from those un-etched area. The
thermal light source at plane $X$ is simulated by
%pseudo-thermal source prepared by
projecting a beam of laser pulse (frequency doubled Nd: YAG impulse
laser,$\lambda=0.532\mu m$) onto a slow-round ground glass disk. the
diffracted dynamic speckle field features the full statistical
character of real thermal light but with much longer coherent time
to allow detecting its intensity fluctuations. Away from thermal
light source plane $X$ $40mm$, we place a beam splitter to form two
optical paths for imaging system referring to Fig.\ref{fig:setup}
shows. The pure phase object $\mathrm{t}(x_{0})$ is placed on plane
$X_{0}$ away from source plane $X$ by $d_{1}=60mm$. Charge Coupled
Device camera CCD-1 is placed at plane $X_{1}$ in the optical path
after the object by $d_{2}=75mm$. Correspondingly, we place CCD-2 at
plane $X_{2}$ on the other optical path, arrange the distance from
the thermal light source $X$ to $X_{2}$ to be $d=135mm$. The
parameters are thus arranged to meet with the condition
$d=d_{1}+d_{2}$, which was required by the theoretical prediction in
Ref.\cite{ch}. The light spot diameter on ground glass is chosen as
$\sigma=3mm$ in order to ensure linear dimensions of the coherence
area\cite{speckle} of the pseudo-thermal light field across the
object plane $D\approx \lambda d_{1}/\sigma=0.532\times60/3=10.64\mu
m$ to be much smaller than the pure object's feature size $150\mu
m$. The exposure time of two CCD cameras is set to $1ms$. Totally
about $10,000$ frames of independent two-dimensional instant
intensity distribution data of speckle fields in the planes of
$X_{1}$ and $X_{2}$ , say, ensembles of $I_{1}(x_{1})$ and
$I_{2}(x_{2})$, are recorded to prepare for correlation operation.
The laser pulse shooting, data acquisition and their recording
process are synchronized and accomplished by computers.

\section{\label{4}Experimental Results}
\begin{figure}
\centerline{\includegraphics* [bb=17 17 150
186,scale=1.2]{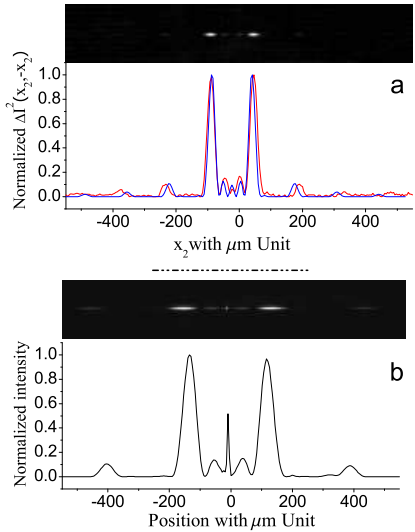}}\caption{\label{fig:giresult}
\textbf{a}. Retrieved two dimensional patterns of $\Delta I(x_{2} ,
-x_{2})$ and its profile (red curve) by using GI schemed setup;
\textbf{b}. The two dimensional Fraunhoufer diffraction pattern and
its profile of the pure phase object obtained by coherent $2f$
system with $f=75mm$ and $\lambda=0.532\mu m$.}
\end{figure}

Choosing  symmetric positions $x_{2}$  and $x_{1}=-x_{2}$  to
calculate $\Delta I^{(2)}(x_{1},x_{2})$ with preserved data
$I_{1}(x_{1})$ and $I_{2}(x_{2})$, we find what we obtain
(Fig.\ref{fig:giresult}a) shares the same pattern in
Fig.(\ref{fig:giresult}b), but have double coordinate scales just as
equation:
\begin{align}
\Delta I ^{2}(x_{2} , -x_{2})=\mid \mathcal F
\{\mathrm{t}(\frac{2\pi[x_{2}-(-x_{2})]}{\lambda
d_{2})}\}\mid^{2}\nonumber\\=\mid \mathcal F \{t(\frac{2\pi
x_{2}}{\frac{\lambda}{2}d_{2}})\}|^{2}, \label{eq:14}
\end{align}
which derived from in Eq.(\ref{eq:12}), anticipated. The result
(Fig.\ref{fig:giresult}a) shows a sub-wavelength interference
pattern equals to the Fraunhoufer diffraction
pattern(Fig.\ref{fig:giresult}b) of the same pure-phase object
accomplished by a $2f$ system(with $\lambda=0.532\mu m$ and
$f=75mm$), but with half of the wavelength of coherent illumination.
In Fig.(\ref{fig:giresult}a), red curve represents profile of two
dimensional experimental result, and blue curve presents the
standard Fourier transform of the pure-phase object's complex
amplitude transmittance under the sub-wavelength condition according
to Eq.(\ref{eq:14}). Its analytical form is given by
\begin{align}
\Delta I ^{2}(x_{2} , -x_{2})=sinc^{2}(\frac{a}{\lambda_{sub}
d_{2}}x_{2})[1-2\nonumber\\
\times\cos(\frac{2\pi a}{\lambda_{sub} d_{2}}x_{2})+4\cos(\frac{4\pi
a}{\lambda_{sub} d_{2}}x_{2})]^{2},\label{eq:15}
\end{align}
in which $a=150\mu m$, $d_{2}=75mm$, and
$\lambda_{sub}=\frac{0.532}{2}\mu m$. From Fig.\ref{fig:giresult}a,
we can see the experiment result fit the theoretical Fourier
transform Eq.(\ref{eq:15}) quite well.

\begin{figure}
\centerline{\includegraphics* [bb=19 17 150
103,scale=1.2]{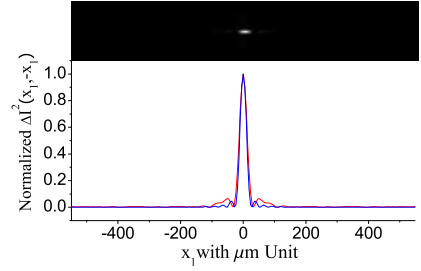}}\caption{\label{fig:hbtresult}
Retrieved two dimensional pattern of $\Delta I(x_{1} , -x_{1})$  and
its profile by using HBT-type of setup, recovers a diffraction
pattern of the limited aperture of the object and retrieves no
information of the phase object. }
\end{figure}

According to Ref.\cite{hbt}, the optical path that passes the object
in our experimental setup is compatible with the classical HBT-type
imaging system if joint detection is accomplished only on the arm
passes the object. So, the 10,000 frames of previously preserved
intensity data recorded on plane $X_{1}$, \emph{i.e.,} ensembles of
$I(x_{1})$ can also be utilized for HBT type of imaging. In this
situation, the two points $x_{1}$ and $x'_{1}$ for joint detecting
are chosen only at plane $X_{1}$ symmetrically with respect to the
center of it, \emph{i.e.}, $x_{1}=-x'_{1}$. The function $\Delta I
^{2}(x_{1} , -x_{1})$ contains no phase information about the object
in this time,  as two-dimensional pattern and its profile (red
curve) in Fig.\ref{fig:hbtresult} shows. Eq.(\ref{13}) gives a
straightforward explanation to the experimental result: The
diffraction pattern we obtained was the Fourier transform of the
square module of the transmitted function of the object. The blue
curve in Fig.{\ref{fig:hbtresult}} describes the Fourier transform
of $|t(x_{0})|^{2}$, with the analytic form of
\begin{align}
\Delta I ^{2}(x_{1} , -x_{1})=sinc^{2}(\frac{5a}{\lambda_{sub}
d_{2}}x_{1}),\label{eq:16}
\end{align} which also fits the experimental result very well.

Thus we find the essential differences of our GI schemed sub-wave
length coherent imaging  from recently reports about sub-wavelength
interference with thermal light \cite{kw1,yshi_epl_2004} is what GI
type of imaging system in our experiment setup retrieved was the
complex amplitude transmittance knowledge of the object rather than
the transmitted intensity as the HBT schemed imaging does.

\section{\label{5}Conclusion and discussion}
The lensless scheme proposes the potential application in hard
$x$-ray, $\gamma$-ray, or other wavelengths where no effective lens
or/and no coherent source is available, and the similar idea have
also been reported brilliantly, but in spatial
domain\cite{apl_yshih}. On the other hand, obtains of sub-wavelength
interference pattern suggest that diffraction limit can be broken
through by using two-photon absorption (TPA) media if we find ways
to fold the symmetric planes of $x_{1}$ and $x_{2}$ into the same
one\cite{dowling}. This feature appeals to quantum lithography which
was widely discussed\cite{gkieee1}.

Apart from the unique sub-wavelength feature of the experiment
result, attention shall also be paid again to the similarity and the
difference between GI schemed and the classical HBT-type of imaging:
They both retrieve diffraction patterns by correlation function
whereas GI recovers knowledge of complex amplitude transmittance
about objects rather than the transmitted intensities as HBT-type
imaging does. Similar conclussion have also been
reported\cite{Lugiato_pra_purephase}, but with use of lens in both
optical paths. In fact our scheme suggests a new way to perform
lensless Fourier transform within fresnel diffraction region.

Unlike the other ghost imaging and ghost diffraction experiments, a
pulsed thermal-like source is used instead of a continuous one, the
intensity correlation can be measured even the exposure time of CCD
camera is much longer than pulse width of the source. This unique
impulse feature of our thermal light source enlightens us on the
issue for recording intensity fluctuation by slow detector. This way
suggests achieving to record intensity fluctuating much faster than
the respond speed of the detecting system. As for hard $x$-ray
imaging, there is a potential applicability to record the
femto-second fluctuations of intensity by using of detecting system
with response speed in nanoseconds.

\begin{acknowledgments}
The authors would like to thank Professor Kaige Wang for helpful
discussion and Professor Yang-chao Tian for preparing the objects.
This research is partly supported by the National Natural Science
Foundation of China, Project No. 60477007, the Shanghai Optical-Tech
Special Project, Project No. 034119815,and Shanghai Dengshan
Project, Project No.60JC14069. \end{acknowledgments}


\begin{thebibliography}{99}

\bibitem{sayre}D. Sayre, Imaging Processes and Coherence in Physics, Springer
Lecture Notes in Physics Vol. 112 (Springer- Verlag, Berlin, 1980),
p. 229.
\bibitem{fienup}J. R. Fienup,Appl.Opt.\textbf{21},2758(1982);
J.Cheng, S. Han, J. Opt. Soc. Am. A \textbf{18}, 1460-1464(2001);
V.Elser, J. Opt. Soc. Am. A. \textbf{20}, 40-55 (2003).
\bibitem{kw1}J. Xiong \emph{et al.}, Phys. Rev. Lett. \textbf{94},
173601 (2005).
\bibitem{yshi_epl_2004}G. Scarcelli, A. Valencia and Y. Shih, Europhys. Lett. \textbf{68},
618(2004).
\bibitem{hbt}R. Hanbury Brown and R. Q. Twiss, Nature (London) \textbf{177}, 27 (1956).
\bibitem{Belinsky}A. V. Belinsky and D. N. Klyshko, Sov. Phys. JETP \textbf{78}, 259
(1994).
\bibitem{cteich_prl_2001}A. F. Abouraddy, B. E. A. Saleh, A. V. Sergienko, and M. C. Teich,
Phys. Rev. Lett. \textbf{87}, 123602 (2001).
\bibitem{boyd_prl_2001}R. S. Bennink, S. J. Bentley, and R. W. Boyd,  Phys. Rev. Lett.
\textbf{89}, 113601 (2002).
\bibitem{lg_prl_2003}A. Gatti, E. Brambilla, and L. A. Lugiato, Phys. Rev.
Lett.\textbf{ 90}.133603 (2003).
\bibitem{boyd_prl_2004}R. S. Bennink, S. J. Bentley, R. W. Boyd, and J. C.Howell, Phys.
Rev. Lett. \textbf{92}, 033601 (2004).
\bibitem{yshi_prl_2004}M. D'Angelo, Y.-H. Kim, S. P. Kulik, and Y. Shih, Phys. Rev.
Lett.\textbf{ 92}, 233601 (2004).
\bibitem{lg_prl_2004}A. Gatti, E. Brambilla, M. Bache, and L. A. Lugiato, Phys. Rev.
Lett. \textbf{93}, 093602 (2004).
\bibitem{ch}J. Cheng and S. Han, Phys. Rev. Lett. \textbf{92}, 093903
(2004).
\bibitem{st_optics}J.W. Goodman, Statistical Optics (Wiley, New York, 1985).
\bibitem{collion}S. A. Collins, J. Opt. Soc. Am. \textbf{60}, 1168 (1970).
\bibitem{yshi_prl_2006}G. Scarcelli, V. Berardi, and Y. Shih, Phys. Rev. Lett. \textbf{96},
063602 (2006).
\bibitem{glauber_pr_1963}R. J. Glauber, Phys. Rev. \textbf{130}, 2529 (1963).
\bibitem{op_coher_qt_op}L.Mandel and E.Wolf, \emph{Optical Coherence and
Quantum Optics}, (Cambridge University Press, New York, 1995),
p.578;
\bibitem{qt_op}D.F.Walls and G.J.Milburn, \emph{Quantum
Optics}(Springer-Verlag,Berlin,1994),p.39.
%\bibitem{q_dtct_theory}D.F.Walls and G.J.Milburn, \emph{Quantum
%Optics}(Springer-Verlag,Berlin,1994),p.39.
\bibitem{speckle}J.W.Goodman, in \emph{Laser speckle and related phenomena}, Laser speckle and related phenomena,edited by J.
C.Dainty (Springer-Verlag, New York, 1984).
\bibitem{apl_yshih}G. Scarcellia, V. Berardi, and Y. Shih, Appl.
Phys. Lett., \textbf{88}, 061106 (2006); Phys. Rev. Lett.,
\textbf{96}, 063602 (2006).
\bibitem{dowling}A. N. Boto \emph{et al.},Phys. \ Rev. \
Lett.\textbf{85}, 2733 (2000).
\bibitem{gkieee1}S.L.Braunstein \emph{et al.}, in Quantum Electronics and Laser
Science Conference, 2001. QELS '01. Techni- cal Digest. (2001), p.
68; M.D¡¯Angelo, M. V. Chekhova, and Y. Shih, Phys. \ Rev. \ Lett.
\textbf{87}, 013602 (2001).
\bibitem{Lugiato_pra_purephase}M. Bache \emph{et al.}, Phys. \ Rev.
\ A, \textbf{73}, 053802 (2006).

\end{thebibliography}
\end{document}